\documentclass[a4paper11pt]{article}

\pdfoutput=1
\usepackage{amsmath}
\usepackage{graphicx}

\usepackage[english]{babel}
\usepackage[utf8]{inputenc}
\usepackage{pdflscape}
\usepackage{enumerate}
\usepackage{amsbsy}
\usepackage{amsmath} 
\usepackage{graphics}
\usepackage{mathrsfs}
\usepackage{wrapfig}
\usepackage{mathtools}

\usepackage{amsfonts}
\usepackage{pstricks}
\usepackage{color}
\usepackage{setspace}

\usepackage{accents}
\usepackage{tensor}

\usepackage{jcappub_ver} 

\newcommand{\bea}{\begin{eqnarray}} \newcommand{\eea}{\end{eqnarray}}
\newcommand{\el}{\nonumber \\}
\newcommand{\re}[1]{(\ref{#1})}

\newcommand{\pat}{\partial}
\newcommand{\abs}[1]{|#1|}

\renewcommand{\a}{\alpha}
\renewcommand{\b}{\beta}
\renewcommand{\c}{\gamma}
\renewcommand{\d}{\delta}
\newcommand{\e}{\epsilon}
\newcommand{\m}{\mu}
\newcommand{\n}{\nu}

\newcommand{\ha}{\frac{1}{2}}

\newcommand{\rmd}{\mathrm{d}}

\title{Inflation with nondynamic distortion to leading order in slow-roll}

\author{Ali Hassan}

\affiliation{University of Helsinki Department of Physics and Helsinki Institute of Physics \\ P.O. Box 64 FIN-00014 University of Helsinki Finland}

\emailAdd{ali.hassan@helsinki.fi}

\abstract{
We study inflation in metric-affine gravity.
We write an action that contains all the second order algebraic distortion terms, and all first order distortion terms with a single covariant derivative, coupled to a scalar field. 
We include the Einstein--Hilbert term with nonminimal coupling, a scalar field potential, and impose projective invariance. 
The distortion equation of motion is algebraic by construction, and the distortion is integrated out analytically. 
This yields a kinetic term sourced entirely by distortion, with a kinetic coupling function determined by the 13 free coupling constants of the starting action. 
We compute inflationary observables for three model classes with a monomial distortion coupling. 
For a monomial potential, the spectral index and tensor-to-scalar ratio depend only on the ratio of the exponents, with the starting coupling constants dropping out entirely; however, this model lies outside the Planck + BK18 $2\sigma$ contours. 
For a potential of the $\alpha$-attractor form, the observables are governed by a single parameter and approach the Starobinsky predictions as a limit. 
Including a nonminimal coupling to the Ricci scalar with a monomial potential can also yield an asymptotically flat effective potential with the same modified Starobinsky observables.
}

\begin{document}

\begin{flushleft}
	\hfill		 HIP-2026-5/TH \\
\end{flushleft}
 
\setcounter{tocdepth}{3}

\setcounter{secnumdepth}{3}

\maketitle

\section{Introduction}

Inflation is the leading framework for describing the early universe \cite{Starobinsky:1979ty, Starobinsky:1980te, Kazanas:1980tx, Guth:1981, Sato:1981, Mukhanov:1981xt, Linde:1981mu, Albrecht:1982wi, Hawking:1981fz, Chibisov:1982nx, Hawking:1982cz, Guth:1982ec, Starobinsky:1982ee, Sasaki:1986hm, Mukhanov:1988jd}, with theoretical predictions that match recent observations remarkably well \cite{Akrami:2018odb, Planck:2018vyg, BICEP:2021xfz, McDonough:2025lzo}. 
The simplest realisation of inflation is driven by a single scalar field, the inflaton, which is generally expected to couple non-minimally to curvature. 
This is due to arguments from effective field theory \cite{Weinberg:2008hq, Solomon:2017nlh}, and as a consequence of explicit loop calculations \cite{Callan:1970ze}). 
Non-minimal couplings of this kind include a direct coupling between a function of the scalar field and the Ricci scalar, as in Higgs inflation \cite{Bezrukov:2007, Bauer:2008}, as well as more general curvature couplings \cite{Capozziello:1999uwa, Capozziello:1999xt, Daniel:2007kk, Sushkov:2009hk, Germani:2010gm, Germani:2010ux, Kobayashi:2010cm, Kamada:2010qe, Kobayashi:2011nu, Germani:2011mx, Tsujikawa:2012mk, Kamada:2012se, Kamada:2013bia, Germani:2014hqa, Yang:2015pga, Kunimitsu:2015faa, DiVita:2015bha, Escriva:2016cwl, Fumagalli:2017cdo, Sato:2017qau, Fu:2019ttf, Granda:2019wyi, Granda:2019wip, Sato:2020ghj, Gialamas:2020vto, Fumagalli:2020ody, Dioguardi:2023jwa, Nezhad:2023dys, BouzariNezhad:2025bgx}.

In the metric formulation, the connection is given by the Levi--Civita connection of the metric, so the metric is the only gravitational degree of freedom. 
In the Palatini formulation (also called the metric-affine formulation), the metric and the connection are independent variables. 
In general both formulations are not equivalent, except if the action is simply the Einstein--Hilbert term with minimally coupled matter \cite{Bauer:2008, Bauer:2010, Tenkanen:2017jih, Annala:2022gtl}. 
Inflation in the Palatini formulation of gravity has been the subject of recent interest \cite{Annala:2021zdt, Rasanen:2017, Rasanen:2018b, Rasanen:2022ijc, Racioppi:2025igu, Enckell:2018a, Tenkanen:2020dge, Dioguardi:2023jwa, Kallosh:2013hoa,Kallosh:2013yoa}. 
The difference between the independent connection and the Levi--Civita connection is the distortion tensor, which encodes the manifold's non-metricity and torsion. 
The distortion tensor vanishes dynamically in the aforementioned Einstein--Hilbert action (except for a projective mode), but in general has a nontrivial solution that influences the physics, and can be treated as an independent dynamical quantity \cite{Borunda:2008, Percacci:2020bzf, BeltranJimenez:2019esp}.

In the case of coupling of the distortion to a function of the scalar field (as in nonminimal coupling to curvature), solving for the distortion and substituting it back into the action can yield an effective kinetic term \cite{Langvik:2020nrs, Hassan:2024vjg, Hassan:2026meg, Annala:2021zdt}.
In the present paper we pursue this idea in an unconstrained metric-affine setting.
We augment the Einstein--Hilbert action in the Palatini formulation with an indepenedent distortion contribution coupled to a function of the scalar field, such that the connection equation of motion is algebraic in the distortion to leading order in the slow--roll regime. 
Hence, the distortion can be integrated out analytically.
We do not include a kinetic term a priori for the scalar field.
We show that, upon integrating out the distortion, the result is an effective scalar field theory in which the entire inflaton kinetic term is sourced by the distortion, with the kinetic coupling function $K(\varphi)$ determined by the starting coupling constants. 
Hence, despite starting from a theory where the scalar field has no kinetic term, it becomes dynamical and can be stable for a wide range of parameters.

We then analyse the inflationary observables for three classes of models, all with a monomial form for the distortion coupling function. 
For a monomial potential, we find that the initial coupling constants drop out of the calculation for the spectral index and tensor-to-scalar ratio, which are instead governed only by a single parameter given by the ratio of the monomial exponents. 
Hence, regardless of the starting constants, the inflationary predictions remain the same.
For a potential of the $\alpha$-attractor form, the observables depend on a single parameter, which weakly shifts the predictions from the Starobinsky limit, with the Starobinsky results still recovered in the limit of this parameter going to zero. 
Therefore, this model's predictions are well within the current Planck~+~BK18 bounds \cite{Bhattacharya:2022akq, Planck:2018vyg, BICEP:2021xfz}.
Including a non-minimal coupling to the Einstein--Hilbert term of the Higgs inflation type generates the same modified-Starobinsky relation but with a transformation of the model parameter \cite{Bezrukov:2007}.

The paper is structured as follows. 
In section~2 we define the geometric quantities, write down the action, then solve for the distortion and insert it back into the starting action. 
Thus we derive the form of the nonminimal coupling function and analyse its stability conditions and discuss special cases including the conditions for dynamical vanishing of torsion or non-metricity.
In section~3 we compute inflationary observables for the aforementioned three model classes and compare with observations. 
We conclude in section~4.

\section{Kinetic term sourced purely from distortion}

\subsection{Definitions of geometric quantities}

In the Palatini formulation the metric $g_{\a\b}$ and the connection $\Gamma^\c_{\a\b}$ are independent variables. We decompose the connection, defined with the covariant derivative as $\nabla_\b A^\a=\pat_\b A^\a + \Gamma^\a_{\b\c} A^\c$, $\nabla_\b A_\a=\pat_\b A_\a - \Gamma^\c_{\b\a} A_\c$, as
\bea \label{Gamma}
  \Gamma^\c_{\a\b} &=& \mathring\Gamma^\c_{\a\b} + L^\c{}_{\a\b} = \mathring\Gamma^\c_{\a\b} + J^\c{}_{\a\b} + K^\c{}_{\a\b} \ ,
\eea
where $\mathring\Gamma_{\a\b}^\c$ is the Levi--Civita connection of the metric $g_{\a\b}$. We denote quantities defined with the Levi--Civita connection by $\mathring{}$. In the second equality we have decomposed the distortion tensor as $L^\c{}_{\a\b}$ into the disformation $J_{\a\b\c}$ and the contortion $K_{\a\b\c}$ as $L^\c{}_{\a\b} = J^\c{}_{\a\b} + K^\c{}_{\a\b}$. 
The disformation and contortion are defined as
\bea \label{JK}
  J_{\a\b\c} &\equiv& \frac{1}{2} \left(Q_{\a\b\c}  - Q_{\c\a\b} - Q_{\b\a\c} \right) \ , \qquad K_{\a\b\c} \equiv \ha (T_{\a\b\c} + T_{\c\a\b} + T_{\b\a\c} ) \ ,
\eea
where $Q_{\a\b\c}$ and $T_{\a\b\c}$ are the non-metricity and the torsion, respectively, defined as
 \bea \label{TQ}
  \qquad Q_{\c\a\b} \equiv \nabla_\c g_{\a\b} = - 2 L_{(\a|\c|\b)} \ , \qquad T^\c{}_{\a\b} &\equiv& 2 \Gamma^{\c}_{[\a\b]} = 2 L^\c{}_{[\a\b]} \ .
\eea
We have $Q_{\c\a\b}=Q_{\c(\a\b)}$, $J_{\a\b\c}=J_{\a(\b\c)}$, and $K^\c{}_\a{}^\b=K^{[\c}{}_\a{}^{\b]}$. 
The Riemann tensor can be decomposed into the Levi--Civita part and the distortion part as
\bea \label{Riemann}
  R^{\a}{}_{\b\c\d} &\equiv& \pat_\c \Gamma^{\a}_{\d\b}-\pat_\d\Gamma^{\a}_{\c\b} + \Gamma^{\a}_{\c\mu} \Gamma^{\mu}_{\d\b} - \Gamma^{\a}_{\d\mu}\Gamma^{\mu}_{\c\b}= \mathring R^{\a}{}_{\b\c\d} + 2 \mathring \nabla_{[\c} L^\a{}_{\d]\b} + 2 L^\a{}_{[\c|\mu|} L^\mu{}_{\d]\b} \ .
\eea
The Ricci tensor is $R_{\a\b}\equiv R^{\c}{}_{\a\c\b}$, and the Ricci scalar is $R\equiv g^{\a\b}R_{\a\b}$.
We define the projective transformation of the connection as 
\begin{equation} \label{projective}
	\Gamma^\c_{\a\b} \to \Gamma^\c_{\a\b} + \delta^\c{}_\b A_\a \ ,
\end{equation}
where $A_\a$ is an arbitrary vector.

\subsection{The action and equations of motion}

Consider the metric-affine action 

\begin{equation} \label{A0}
    S = \int \rmd^4x \sqrt{-g} \Big\{ \ha F(\varphi) R
    + P(\varphi) \mathcal M (\nabla L, L^2)
    - V(\varphi) \Big\} \ ,
\end{equation}
where $R$ is the Ricci scalar in the metric-affine formulation. 
Here $\mathcal{M}$ encompasses all independent contractions of the distortion tensor up to quadratic order and all total derivative terms linear in $L_{\alpha\beta\gamma}$. It is given by
\begin{align} \label{M} 
\mathcal M = &
\nabla_\a (a_1 L^{\a \m}{}_\m + a_2 L^\m{}_\m{}^\a + a_3 L^{\m \a}{}_\m)
\\ & \nonumber
+ b_1 \tensor{L}{^\a ^\b _\b} \tensor{L}{_\a ^\c _\c} 
+ b_2 \tensor{L}{_\a _\b _\c} \tensor{L}{^\a ^\b ^\c} 
+ b_3 \tensor{L}{_\a _\c _\b} \tensor{L}{^\a ^\b ^\c} 
+ b_4 \tensor{L}{^\a ^\b ^\c} \tensor{L}{_\b _\a _\c} 
+ b_5 \tensor{L}{^\a ^\b ^\c} \tensor{L}{_\b _\c _\a} 
\\ & \nonumber
+ b_6 \tensor{L}{^\a _\a ^\b} \tensor{L}{^\c _\b _\c} 
+ b_7 \tensor{L}{^\a _\b _\a} \tensor{L}{^\c _\b _\c} 
+ b_8 \tensor{L}{^\a ^\b ^\c} \tensor{L}{_\c _\b _\a} 
+ b_9 \tensor{L}{^\a _\a ^\b} \tensor{L}{^\c _\b _\c}
+ b_{10} \tensor{L}{^\a _\b _\a} \tensor{L}{^\c _\b _\c} 
+ b_{11} \tensor{L}{^\a _\a ^\b} \tensor{L}{^\c _\c _\b} \ ,
\end{align}
where we defined 14 constants $a_i$ and $b_i$, each coupled respectively with one of the 14 independent terms. 

In general, the most comprehensive algebraic action would also include higher order distortion terms.
But such terms would be higher-order in the slow-roll approximations given the form of the solution for the distortion, as we will show in this section.
Therefore it suffices to stop after the second order distortion contribution.

On the other hand, adding more covariant derivatives would render the distortion dynamical (unless if the coupling constants are chosen such that the dynamical terms cancel in the equation of motion), which is in general not possible to integrate out analytically. 
However, we note that higher order covariant derivatives will either result in higher-order time derivatives, the effects of which would be suppressed under the quasi-static time evolution of slow-roll inflation, or result in a contribution of extra powers of the Hubble parameter $H$ to the coupling function $P$, such that we would have $P = P(\varphi,H, \dot H)$.
In fact, one could also consider the direct coupling of distortion to curvature. 
On an FLRW background this would carry the same effect of allowing the coupling function $P(\varphi)$ to also depend on the Hubble parameter $H$.

However, in analysis of slow-roll inflation, the time evolution of the Hubble parameter is slow-roll suppressed with respect to the Hubble parameter itself. 
And the Hubble parameter terms would be replaced by the slow-roll equation $3 H^2 = V(\varphi)$. 
Hence, for the purpose of calculating inflationary observables to leading order in slow-roll, admitting dependence on curvature would effectively only alter the coupling function $P(\varphi)$.

If we choose that the action respects projective invariance then we would have the freedom to set one of the $14$ constants to zero. Therefore we henceforth make the choice $a_3 \rightarrow 0$, and we are left with 13 coupling constants that parametrise our starting action. 

Finally, we note that in the cases of vanishing torsion we would have 7 terms: 2 arising from one power of the disformation tensor $J_{\a\b\c}$ and 5 terms from the disformation-squared contribution. As for vanishing nonmetricity we would have~4, of which only one is from the covariant derivative of one power of the torsion tensor. We note that the sum of the two cases individually is not 14 as the full unconstrained case, since 3 terms come from cross terms between torsion and nonmetricity (2 of which are cross between the traces).

Proceeding, we perform integration by parts to write the action as
\begin{align} \label{A1}
    S = & \int \rmd^4x \sqrt{-g} \Big\{ \ha F(\varphi) R - V(\varphi) 
- \partial_\a P (a_1 L^{\a \m}{}_\m + a_2 L^\m{}_\m{}^\a )
\\ & \nonumber
+ P ( b_1 \tensor{L}{^\a ^\b _\b} \tensor{L}{_\a ^\c _\c} 
+ b_2 \tensor{L}{_\a _\b _\c} \tensor{L}{^\a ^\b ^\c} 
+ b_3 \tensor{L}{_\a _\c _\b} \tensor{L}{^\a ^\b ^\c} 
+ b_4 \tensor{L}{^\a ^\b ^\c} \tensor{L}{_\b _\a _\c} 
+ b_5 \tensor{L}{^\a ^\b ^\c} \tensor{L}{_\b _\c _\a} 
\\ & \nonumber
+ b_6 \tensor{L}{^\a _\a ^\b} \tensor{L}{^\c _\b _\c} 
+ b_7 \tensor{L}{^\a _\b _\a} \tensor{L}{^\c _\b _\c} 
+ b_8 \tensor{L}{^\a ^\b ^\c} \tensor{L}{_\c _\b _\a} 
+ b_9 \tensor{L}{^\a _\a ^\b} \tensor{L}{^\c _\b _\c}
+ b_{10} \tensor{L}{^\a _\b _\a} \tensor{L}{^\c _\b _\c} 
+ b_{11} \tensor{L}{^\a _\a ^\b} \tensor{L}{^\c _\c _\b} )
    \Big\} \ ,
\end{align}
where we have dropped a boundary term. At this stage we perform a conformal transformation given by $g_{\a\b}\to F^{-1} g_{\a\b}$ to set the coupling function in front of the Ricci scalar to unity. This will correspond to shifting the functions potential as $V\to V/F^2$, $P \to P / F$ and $P' \to P' / F$.

Upon taking the variation of the action (\ref{A1}) with respect to the distortion tensor $L_{\a\b\c}$, we retrieve the following equation of motion
\begin{align} & \label{LEM}
    P \Big( 2 b_1 \tensor{L}{_\a ^\m _\m} g_{\b\c} 
    + 2 b_2 \tensor{L}{_\a _\b _\c} 
    + 2 b_3 \tensor{L}{_\a _\c _\b} 
    + 2 b_4 \tensor{L}{_\b _\a _\c} 
    + b_5 ( \tensor{L}{_\b _\c _\a} + \tensor{L}{_\c _\a _\b} ) 
    + b_6 ( g_{\a\b} \tensor{L}{_\c ^\m _\m} + \tensor{L}{^\m _\m _\a} g_{\b\c} )
\nonumber \\ & \nonumber
    + b_7 ( g_{\a\c} \tensor{L}{_\b ^\m _\m} + \tensor{L}{^\m _\a _\m} g_{\b\c} )
    + 2 b_8 \tensor{L}{_\c _\b _\a} 
    + b_9 ( g_{\a\b} \tensor{L}{^\m _\c _\m} + g_{\a\c} \tensor{L}{^\m _\m _\b} ) 
    + 2 b_{10} g_{\a\c} \tensor{L}{^\m _\b _\m} 
    + 2 b_{11} g_{\a\b} \tensor{L}{^\m _\m _\c}  \Big)
\\ &
 - P' (a_1 g_{\b\c} \partial_\a \varphi 
    + a_2 g_{\a\b}\partial_\c \varphi ) 
    + \ha g_{\a\b} \tensor{L}{_\c ^\m _\m} 
    + \ha \tensor{L}{^\m _\m _\a} g_{\b\c} 
    - \ha \tensor{L}{_\b _\c _\a} 
    - \ha \tensor{L}{_\c _\a _\b} = 0 \ .
\end{align}
And now to solve the equation of motion (\ref{LEM}) we employ the following ansatz as the general solution for torsion and nonmetricity sourced by the gradient of a scalar field
\begin{align}
    & Q_{\a\b\c} = P' ( 2 q_1 g_{\a(\b} \partial_{\c)} \varphi + q_2 g_{\b\c} \partial_\a \varphi )
\\ &
    T_{\a\b\c} = 2 P' t_1 g_{\a [ \b} \partial_{\c ] } \varphi \ .
\end{align}
Where $q_i$ and $t_1$ are constants to be determined from the equation of motion (\ref{LEM}). 
Solving for the torsion and nonmetricity one retrieves the following solution for the distortion tensor
\begin{equation} \label{Lsol}
    L_{\a\b\c} = 
- \frac { P' \big( P (c_1 + c_2 P) g_{\b\c} \partial_\a \varphi 
+ (c_3 + P (c_4 + c_5 P)) g_{\a\c}\partial_\b \varphi 
+ P (c_6 + c_7 P) g_{\a\b}\partial_\c \varphi \big) } {P (c_8 + P (c_{9} + c_{10} P) )} \ 
\end{equation}
where we have defined the ten new constants $c_i$, which are given by some rather involved combinations of the constants $a_i$ and $b_i$. We give these definitions in Appendix \ref{App1}.

\subsection{Special cases}

Here a few comments are in order. First, we note that it is always possible, through a judicious choice of initial coupling constants, to get the torsion or nonmetricity to vanish dynamically (as a solution of the equations of motion).
This reflects some symmetry of the starting action and is in general not equivalent to starting with a theory that assumes the vanishing of the geometric quantity a priori. 
One example realisation of this would be for any starting action that respects the following 3 constraints 
\begin{align} &
b_{1} = 4 b_{10} + 2 b_{11} + b_{2} + b_{3} + \ha b_{5} + \ha b_{6} + b_{8} + 3 b_{9}
\\ & 
b_4 = 4 b_{10} + b_2 + b_3 - b_7 + b_8 + 2 b_9
\\ & 
a_1 = - a_2 \ , 
\end{align}
which would be of dynamically vanishing torsion and only propagate nonmetricity degrees of freedom.

Furthermore, we note that the system does not allow for any dynamic solution wherein both the torsion and nonmetricity vanish simultaneously through a deliberate choice of starting constants $a_i$ and $b_i$, except for the trivial solution $a_1 = a_2 = 0$, where the source term is completely removed from the equation of motion (\ref{LEM}). 
So the entire distortion vanishing dynamically, as is the case of the Einstein--Hilbert action in the Palatini formalism, is not allowed here. 
It would however be possible if the coupling constants $b_i$ where promoted to be functions of the field $\varphi$ instead, such that $b_i(P^{-1}(\varphi))$, in order to allow some cancellations with the terms arising from the Einstein--Hilbert contribution.
In the case of a nontrivial Einstein--Hilbert coupling function $F$, that would correspond to the constraint $P = F$.

\subsection{The effective action}

We proceed by inserting the solution for the distortion tensor $L_{\a\b\c}$ given by equation (\ref{Lsol}) back into the action (\ref{A1}), which yields the following effective action
\begin{equation}
    S = \int \rmd^4x \sqrt{-g} \Big\{ \ha R 
    - \ha K X
    - V \Big\} \ ,
\end{equation}
with the kinetic coupling function 
\begin{equation} \label{K}
K = - 2
    \frac{P'^2}{P} \frac{d_3 P^2 + d_2 P + d_1}{d_6 P^2 + P  d_5 + d_4} \ .
\end{equation}
Here we have defined new constants $d_i$ (once more we give their relation to the starting constants $a_i$ and $b_i$ in Appendix \ref{App2}.)
%

We note here that in order for the theory to remain stable, this kinetic function must be bounded from below \cite{Hassan:2026meg}.
This forces some restrictions on the constants $d_i$, and hence on the starting constants $a_i$ and $b_i$.  
A sufficient, although not necessary, condition would be a strictly positive choice of kinetic coupling function $K$. 
In that case, for a positive coupling function $P > 0$, we require that the fraction in equation (\ref{K}) be negative. 
This happens in only one of two cases; when the numerator and denominator are of opposite signs at all times. 
More concretely we require 
\begin{align} & \label{Cond1}
    d_3 \leq 0 
    \quad \text{and}   \quad 
    d_2^2 - 4 d_1 d_3 \leq 0 \ 
\\ & \nonumber
    \hspace{-1.27cm} \text{while} \quad 
    d_6 \geq 0
    \quad \text{and}   \quad 
    d_5^2 - 4 d_4 d_6 \geq 0 \ .
\end{align} 
or vice versa, while maintaining that the denominator is nonzero. 

However for a choice of negative coupling function $P$ we require that the fraction remains positive and hence that the numerator and denominator remain of the same sign at all times. Nonetheless we note that as a practical example, conditions (\ref{Cond1}) are automatically satisfied upon setting all the constants $a_i$ and $b_i$ to unity.

In order to proceed more concretely we now give a form for the coupling function $P$ and potential $V$, and then calculate the spectral index and tensor-to-scalar ratio for these setups.

\section{Inflationary observables}

\subsection{Monomial coupling $P$ with $F=1$}

We restrict our discussion in this section to the conditions $d_3 \neq 0$ and $d_6 \neq 0$. 
For analysis of large field inflation we take the large field limit of (\ref{K}) for a monomial coupling function $P = P_0 (\varphi - \n)^p \simeq P_0 \varphi^p$, where $\nu$ is a constant introduced in order to avoid the limit $P=0$ at the end of inflation. 
In the large field limit, for $p \geq 2$, the kinetic coupling function becomes 
\begin{equation}
K 
\approx 
- 2 p^2 k_0 \varphi^{p - 2} 
\ ,
     \quad \text{where} \quad 
     k_0 \equiv \left|P_0 \frac{d_3}{d_6}\right| \ .
\end{equation}
First we note that we require the combination of constants $P_0 \frac{d_3}{d_6}$ to be negative, as otherwise the kinetic coupling function will not be bounded from below in the large field limit. 
Therefore we would have the canonical scalar field $\chi$ given by 
\begin{equation}
    \chi = \int\rmd\varphi \sqrt{K} \simeq
    2 \sqrt{2 k_0} \varphi^{\frac{p}{2}} 
     \ .
\end{equation}
And therefore $\varphi \propto \chi^{\frac{2}{p}}$.

\subsubsection{Monomial potential}

Given a simple monomial potential of the form $V = V_0 \varphi^n$ we have 
\bea
  V &\propto& \chi^{\frac{2 n}{p}} \ .
\eea
Such a potential gives the slow-roll parameters
\bea \label{SR}
  \epsilon &\equiv& \ha \left( \frac{V'}{V} \right)^2 
  = \frac{2 n^2}{p^2} \chi^{-2} \el
  \eta &\equiv& \frac{V''}{V} 
  = - \frac{2 n (p - 2 n) }{p^2} \chi^{-2} \ ,
\eea
for which the spectral index is
\bea \label{ns1a}
  n_s &=& 1 - 6 \epsilon + 2 \eta 
  = 1 - \frac{4 n ( p + n )}{p^2} \chi^{-2} \ .
\eea
The number of e-folds is
\bea \label{N1}
  N &=& \int_{\chi_\textrm{end}}^\chi \frac{\rmd\chi}{\sqrt{2\epsilon}} = \frac{p}{4 n} \chi^2 \ 
\eea
where we have assumed $\chi \gg \chi_\textrm{end}$; the subscript refers to the end of inflation. Inserting $\chi$ in terms of $N$ from \re{N1} into \re{ns1a} we obtain
\begin{equation} \label{ns1b}
  n_s = 1 - \frac{1}{N} ( 1 + \frac{n}{p}) \ , \quad 
 r = \frac{8 n}{N p} \ ,
\end{equation}
for which the only parameter is the ratio $n/p$. 
For such a model $n=p$ yields $n_s = 0.96$--$0.967$ for $N=50$--$60$ respectively, while $n = 2 p$ yields $n_s = 0.94$--$0.95$. 
Whereas the tensor-to-scalar ratio is in general too large as to be ruled out by the Planck + BK18 observational upper-limit $r < 0.036$. 
For $N=50$, we have $r=0.16 n/p$, which would not be small enough unless we go to the limit of $\frac{n}{p} \leq 0.2$, which would violate then the observational limits on $n_s$. 

\begin{figure}
    \centering
    \includegraphics[width=1\linewidth]{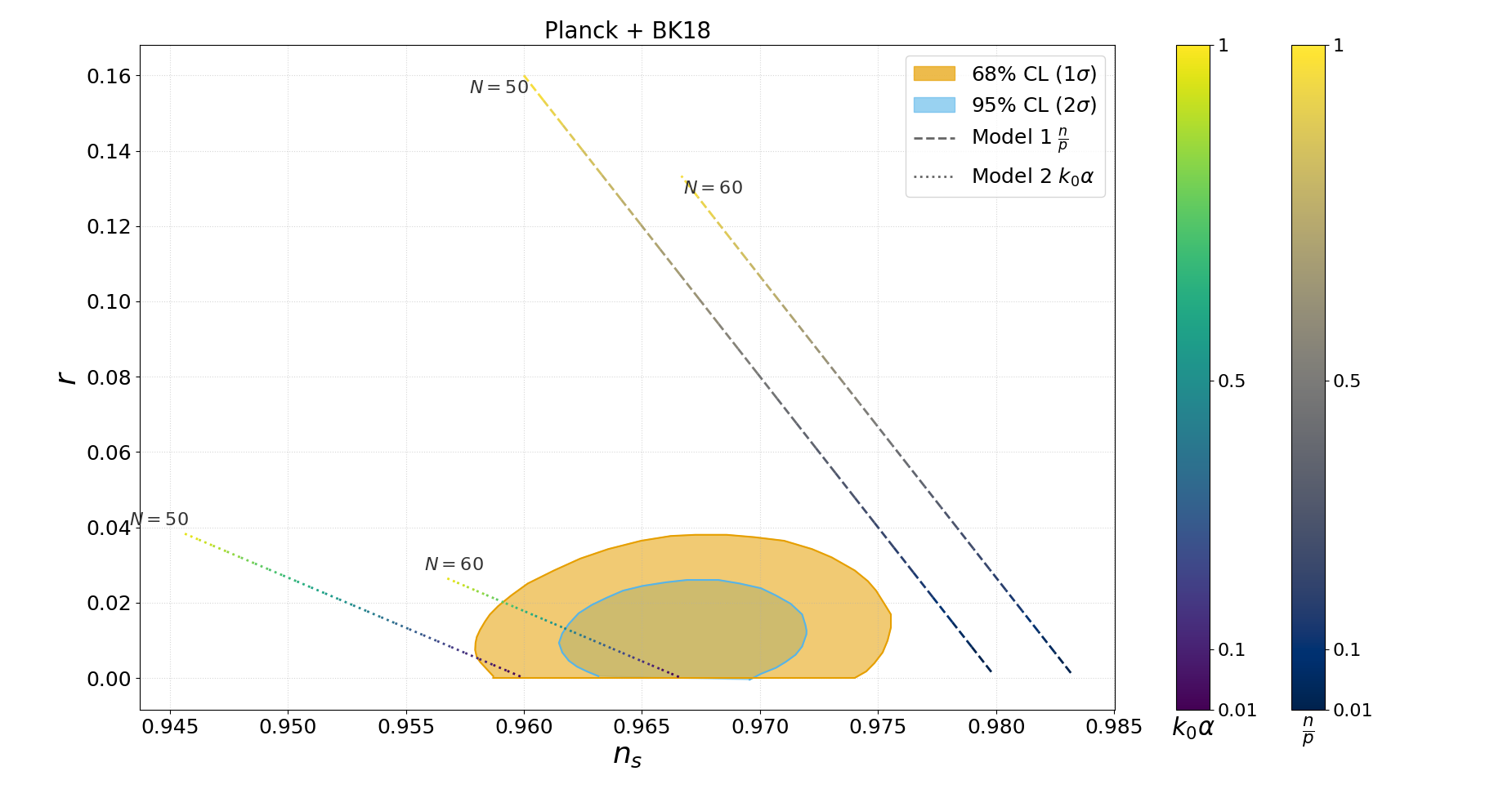}
\caption{
Inflationary observables in the $n_s$-$r$ plane, compared against the Planck + BK18 observational constraints \cite{Akrami:2018odb, Planck:2018vyg, BICEP:2021xfz}. 
The shaded orange and blue regions denote the $68\% (1\sigma)$ and $95\% (2\sigma)$ confidence level contours respectively. 
Three models are shown for monomial coupling $P=P_0 \varphi^p$, and for both $N=50$ and $N=60$ e-folds. 
Model 1 (dashed lines) of \eqref{ns1b} corresponds to a monomial potential $V = V_0 \varphi^n$, parametrised only by the ratio of exponents $n/p$ ranging from 0.01 to 1.
Model 2 (dotted lines) of \eqref{ns2b} corresponds to an $\alpha$-attractor potential, parametrised by $k_0 \alpha$ ranging from $0.01$ to $1$; this model enters the observational contours in the limit of small $k_0\alpha$, approaching the Starobinsky prediction. 
The colour scale on the right indicates the parameter values for each model.}
\label{Fig1}
\end{figure}

In figure \ref{Fig1}, we plot different values of $\frac{n}{p}$ between $\frac{n}{p}=1$ and $\frac{n}{p}=0.01$ in terms of the spectral index $n_s$ against the tensor-to-scalar ratio $r$ with a dashed line for both values of $N=50$ and $N=60$, labeled 'model 1'. 
The Planck 2018 + BK18 constraints are also plotted in the same figure for comparison \cite{Planck:2018vyg, BICEP:2021xfz}.
This model lies outside the $2 \sigma$ range of the observables $n_s$ and $r$. 

However, we note that taking to account DESI BAO data may push the value of $n_s$ upwards, which may make this model more viable for $N=50$, and especially if we also consider values for $N$ slightly below that \cite{Balkenhol:2025wms,AtacamaCosmologyTelescope:2025blo,Ferreira:2025lrd}.
Moreover, a more careful accounting of the end of inflation and preheating might slightly alter these values.

It is interesting that the starting constants do not contribute at this order to the observables, and that the only contribution comes from the single parameter $n/p$. 
Hence these inflationary observables characterise a large set of starting actions, irrespective by the different possible choices of the 13 starting coupling constants.

\subsubsection{Potential of the $\a$-attractor type}

For the same coupling function $P = P_0 \varphi^p$, we instead consider a potential of the $\a$-attractor form \cite{Kallosh:2013yoa,Kallosh:2013hoa}
\begin{equation} \label{Vstar}
    V = V_0 ( 1 - e^{- \sqrt{\frac{2}{3\a}} \varphi})^{2n}
    = V_0 ( 1 - e^{- \kappa \chi^{\frac{2}{p}} })^{2n}
    \ , 
\end{equation}
which reproduces the Starobinsky limit $\a = 1$, $n=1$ \cite{Bhattacharya:2022akq, Mukhanov:1981xt,Starobinsky:1980te, Whitt:1984pd}. 
The constant $\kappa \equiv \sqrt{\frac{2}{3 \a}} (2^3 k_0)^{-\frac{1}{p}} $ is defined for convenience. 
The leading-order contributions to $\e$ and $\eta$ become
\begin{align} &
\e \simeq  
8 \frac{n^2 \kappa^2}{p^2} e^{ - 2 \kappa \chi^{ \frac{2}{p}} } \chi^{ \frac{4}{p} - 2 }
\\ &
\eta \simeq 
- 4 \frac{n \kappa}{p^2} \chi^{ \frac{2}{p} - 2 } e^{ - \kappa \chi^{ \frac{2}{p}} } 
( - 2 + p + 2 \kappa \chi^{\frac{2}{p}} )\ , 
\end{align}
For the case $p=2$ we have the factor $\chi^{\frac{4}{p}-2} = 1$ from the slow--roll parameters $\e$ and $\eta$, and we end up with a  modification of the well-known Starobinsky results for $n_s$ and $r$ as 
\begin{equation} \label{ns2b}
    n_s = 1 - \frac{2}{N} - \frac{36 k_0 \a}{N^2}
    \quad ,  \quad  
    r = \frac{96 k_0 \a}{ N^2} \ .
\end{equation}
Here once more the inflationary model is described by only one parameter, given by the multiple $k_0 \a = \a \abs{P_0 \frac{d_3}{d_6}}$. Remarkably, these results do not depend on the power $n$. Hence we have shown that for $p=2$, if the potential is to start out as $\a$-attractor type, we would end up with the same inflationary observables - but with the added generalisation that it is now tunable by the ratio of starting constants $d_3$ and $d_6$ and coupling constant $P_0$, regardless of the starting power $n$ of the potential. 

For the limit $k_0 \alpha \to 0$, we have $(n_s,r)\to(1 - \frac{2}{N},0)$, which reproduces Starobinsky results for $n_s$ while further suppressing $r$ \cite{Martin:2013tda}. These results are plotted in figure \ref{Fig1} with a dotted linebar for $N=50$ and $N=60$ and $k_0 \alpha$ goes from $k_0 \alpha=1$ to $k_0 \alpha=0.01$. It is labeled as "model 2".



\subsection{Nonminimal coupling to Einstein--Hilbert}

If we include the nonminimal coupling of the Einstein--Hilbert contribution $F R$, we effectively shift $V\to V/F^2$ while $P \to P / F$ and $P' \rightarrow P'/F$. 
This means that in the Einstein frame the kinetic coupling function now becomes
\begin{equation}
    K =
    - 2 \frac{P'^2}{P F}
    \frac{d_3 P^2 + d_2 FP + d_1 F^2}{d_6 P^2 + d_5 FP + d_4 F^2} \ .
\end{equation}
We again take the same monomial form for $P=P_0 \varphi^p$. As for $F$, we make the choice $F = 1 + \xi \varphi^q$ inspired by the Higgs inflation model \cite{Bezrukov:2007} where $q=2$. 
We take the terms to be of the same dimension, so we henceforth set $q = p$.
The leading order contribution becomes 
\begin{equation} \label{betaDef}
    K \simeq 
    p^2 \b^2 \varphi^{-2}
    \ \ , \ 
    \text{where} \ \ 
    \b^2 = 2 \left|\frac{P_0}{\xi} \ \frac{d_1 \xi^2 + P_0 d_2 \xi + d_3 P_0^2}{d_4 \xi^2 + P_0 d_5 \xi + d_6 P_0^2 }\right|
    \ ,
\end{equation}
and therefore we have 
$\varphi = e^{\chi / \beta}$.
We now define the effective potential, given by the conformal relation $U(\varphi) \equiv V/F^2 = V(1 + \xi \varphi^2)^{-2}$. 

\subsubsection{Monomial potential}

For a monomial potential $V = V_0 \varphi^n$, the effective potential takes the form $U = V_0 \varphi^n (1 + \xi \varphi^2)^{-2}$. 
In terms of the canonical field $\chi$, the effective potential reads
\begin{equation}
    U = 
    V_0 e^{n \chi / \b} (1 + \xi e^{p \chi / \b} )^{-2} \ .
\end{equation}
We note that for values of $n=2p$, the potential is asymptotically flat. 
This for example would match the Higgs inflation with quartic potential, where $p=2$ and $n=4$ \cite{Bezrukov:2007}. 
Let us investigate the slow--roll behaviour of such asymptotically flat models.

The slow--roll parameters are
\begin{align} &
    \e = 
\frac{2 p^2}{\xi^2 \b^2} e^{-2 p \chi / \b} 
\\ &
\eta = 
- \frac{2 p^2}{\xi \b^2} e^{- p \chi / \b} 
\ .
\end{align}
%
%
And hence, the spectral index and tensor-to-scalar ratio become 
\begin{align}
    n_s = 1 - \frac{2}{N} - \frac{3 \b^2}{ N^2 p^2 }
\quad , \quad 
r = \frac{8 \b^2}{ N^2 p^2 } 
 \ .
\end{align}

This returns the same modified Starobinsky relations as model 2, and hence would completely coincide it on the $n_s$ -- $r$ plane, only rescaled for different values of the model parameter $\frac{\b^2}{p^2}$. 
Here too Starobinsky results are retrieved for the model parameter limit $\frac{\b^2}{p^2} \to 0$.
We further note that a more careful accounting of the end of inflation may shift these values slightly, as is the case with all of the analyses in this paper.  

On the other hand, if we were to choose the Starobinsky potential and $\a$-attractor models (or any other asymptotically flat potential), while keeping these same choices for $F$ and $P$, we would not end up with an effective potential $U = V_0 \varphi^n(1 + \xi \varphi^2)^{-2}$ that yields an inflationary behaviour. 
That is because, for a potential function V that is already asymptotically flat, suppressing them with the factor $1/F^2$ with a growing function $F$ (for example being of the Higgs inflation form), would yield an overall effective potential that asymptotes to zero.

\pagebreak

\section{Conclusion}

We have considered inflation in the metric-affine formulation of gravity, where the metric and connection are independent variables \cite{Bauer:2008}.
We wrote down a distortion contribution that consists of all the terms quadratic and algebraic in distortion, and all linear terms in distortion with a single covariant derivative, and coupled it to a function of the scalar field.
This contribution of distortion keeps its equation of motion algebraic to leading order in the slow-roll approximations.
The action is parametrised by 14 coupling constants, which we drop to 13 after imposing projective  invariance. 

We solved the distortion equation of motion using an ansatz and inserted the solution back into the action to obtain an effective scalar--tensor theory in which the inflaton acquires a kinetic term sourced entirely by the distortion. 
The kinetic coupling function $K(\varphi)$ is determined by the starting coupling constants through equation (\ref{K}), and the scalar field dynamics are stable for 2 separate sets of 4 independent constraints on the 13 parameter choices.
We also note that it is always possible, through a judicious choice of starting constants, to have either the torsion or the nonmetricity vanish dynamically (although not both at the same time), without having to impose this as a constraint on the geometry a priori. 

We then computed the inflationary observables for three model classes with a monomial coupling function $P = P_0 \varphi^p$.
For a monomial potential $V = V_0 \varphi^n$, the spectral index and tensor-to-scalar ratio are governed solely by the ratio $n/p$, with the 13 starting coupling constants dropping out entirely at leading order.
This demonstrates the robustness of the model's predictions across a large class of starting actions. 
However, this model lies outside the $2\sigma$ observational contours of Planck~+~BK18 for all values of $n/p$, although it may be more consistent with the DESI BAO data \cite{Balkenhol:2025wms,AtacamaCosmologyTelescope:2025blo,Ferreira:2025lrd,Akrami:2018odb,Planck:2018vyg,BICEP:2021xfz}.
For a potential of the $\alpha$-attractor form with $p=2$, the observables are controlled by the single parameter $k_0\alpha$, which generalises the Starobinsky predictions \cite{Bhattacharya:2022akq, Starobinsky:1980te, Martin:2013tda}. 
This model enters the $1\sigma$ observational contours in the small $k_0\alpha$ regime and is insensitive to the power $n$ of the potential.
Including a nonminimal coupling to the Einstein--Hilbert term of the form $F = 1 + \xi\varphi^p$ with a monomial potential satisfying $n=2p$, generates an asymptotically flat effective potential \cite{Bezrukov:2007, Bauer:2010}. 
In this case, the observables are exactly the same as the previous model, up to a redefinition of the model parameter, which is now given by \eqref{betaDef}.
A careful consideration of reheating may shift slightly the predicted values of $n_s$ and $r$. 
These results also depend on the large field approximation. 

We delegate to future work the inclusion of dynamical terms with higher-order covariant derivatives, although these will in general introduce new degrees of freedom, some of which may not be stable.


\section*{Acknowledgments}

AH thanks Syksy Räsänen, Mark Hindmarsh, and Antonio Racioppi for their helpful ideas and fruitful discussions. AH acknowledges support from the Research Council of Finland project 363676.

\newpage


\pagebreak

\appendix

\section{Definition of coupling constants $c_i$} \label{App1}

Below we give the definitions of $c_i$ in \eqref{Lsol}

\begin{align}
  c_{1} = & -a_{2} (2 b_{1} + 24 b_{10} + 6 b_{2} + 2 b_{4} + b_{5} + b_{6} + 7 b_{7} + 6 b_{8} + 3 b_{9}) 
\\ & \nonumber 
+ a_{1} (8 b_{10} + 2 b_{11} + 2 b_{2} + 2 b_{3} + b_{5} + b_{6} + b_{7} + 2 b_{8} + 5 b_{9}) 
\end{align}
\begin{align}
  c_{2} &= -2 \big(a_{2} (-4 b_{3} b_{4} - 2 b_{3} b_{5} - 2 b_{4} b_{5} - b_{5}^2 - 2 b_{3} b_{6} - b_{5} b_{6} - 8 b_{3} b_{7} - 2 b_{4} b_{7} - 4 b_{5} b_{7} 
\\ & \nonumber
+ 3 b_{6} b_{7} - 3 b_{7}^2 + 2 b_{5} b_{8} + 8 b_{6} b_{8} + 4 b_{7} b_{8} + 4 b_{8}^2 + 2 b_{2} (b_{5} + 4 b_{6} + b_{7} + 2 b_{8}) 
\\ & \nonumber
+ b_{10} (-4 b_{4} + 6 b_{5} + 30 b_{6} + 16 b_{8}) + 2 b_{1} (6 b_{10} + 2 b_{2} - 2 b_{3} - b_{5} + 2 b_{8} - 3 b_{9}) 
\\ & \nonumber
- (8 b_{4} + 3 b_{5} + 15 b_{7} - 2 b_{8}) b_{9}) + a_{1} (-4 b_{2}^2 - 4 b_{2} b_{4} + (2 b_{3} + b_{5})^2 - 2 b_{2} b_{6} + 2 b_{3} b_{6} + b_{5} b_{6} 
\\ & \nonumber
- 2 b_{10} (30 b_{11} + 8 b_{2} - 2 b_{3} + 8 b_{4} - b_{5} + 3 b_{6}) - 2 b_{2} b_{7} + 2 b_{3} b_{7} - 2 b_{4} b_{7} + b_{5} b_{7} 
\\ & \nonumber
+ 2 b_{11} (-8 b_{2} + 2 b_{3} + b_{5} - 3 b_{7} - 8 b_{8}) - 4 b_{2} b_{8} - 4 b_{4} b_{8} - 2 b_{6} b_{8} + b_9 (-4 b_{2} + 16 b_{3} 
\\ & \nonumber
- 2 b_{4} + 8 b_{5} + 3 (b_{6} + b_{7}) - 2 b_{8}) + 15 b_{9}^2)\big)
\end{align}
\begin{align}
c_{3} &= -a_{1} - a_{2} 
\end{align}
\begin{align}
c_{4} &= a_{2} (6 b_{1} + 4 b_{10} + 2 b_{2} + 6 b_{4} + b_{5} - 3 b_{6} + 11 b_{7} - 2 b_{8} - b_{9})
\\ & \nonumber
+ a_{1} (4 b_{10} + 6 b_{11} + 2 b_{2} + 6 b_{3} + b_{5} - 3 b_{6} - b_{7} - 2 b_{8} + 11 b_{9}) 
\end{align}
\begin{align}
c_{5} &= -2\big(a_{1} (4 b_{2} b_{4} + 4 b_{4}^2 + 2 b_{2} b_{5} - 2 b_{3} b_{5} + 2 b_{4} b_{5} - b_{5}^2 - 8 b_{3} b_{6} + 2 b_{4} b_{6} - 3 b_{5} b_{6}
\\ & 
+ 8 b_{2} b_{7} + 8 b_{4} b_{7} - b_{5} b_{7} + b_{11} (-4 b_{3} + 16 b_{4} + 6 b_{5} + 30 b_{7}) + 2 b_{10} (6 b_{11} + 2 b_{2} + 2 b_{4} 
\nonumber \\ & 
- b_{5} - 3 b_{6} - 2 b_{8}) - 4 b_{3} b_{8} - 2 b_{5} b_{8} - 2 b_{7} b_{8} + (2 b_{2} - 2 b_{3} + 4 b_{4} - 4 b_{5} - 15 b_{6} + 3 b_{7} 
\nonumber \\ & 
- 8 b_{8}) b_{9} - 3 b_{9}^2) + a_{2} (4 b_{2} b_{3} + 4 b_{3}^2 + 2 b_{2} b_{5} + 2 b_{3} b_{5} - 2 b_{4} b_{5} - b_{5}^2 + 2 b_{3} b_{6} - 8 b_{4} b_{6} 
\nonumber \\ & 
- 3 b_{5} b_{6} + 2 b_{2} b_{7} + 4 b_{3} b_{7} - 2 b_{4} b_{7} - 4 b_{5} b_{7} - 15 b_{6} b_{7} - 3 b_{7}^2 - 4 b_{4} b_{8} - 2 b_{5} b_{8} - 8 b_{7} b_{8} 
\nonumber \\ & 
+ b_{10} (4 b_{2} + 4 b_{3} - 2 (b_{5} + 3 b_{6} + 2 b_{8})) + (8 b_{2} + 8 b_{3} - b_{5} + 3 b_{7} - 2 b_{8}) b_{9} 
\nonumber \\ & \nonumber 
+ 2 b_{1} (6 b_{10} + 8 b_{3} - 2 b_{4} + 3 b_{5} + 15 b_{9}))\big)
\end{align}
\begin{align}
c_{6} &= a_{2} (2 b_{1} + 8 b_{10} + 2 b_{2} + 2 b_{4} + b_{5} + b_{6} + 5 b_{7} + 2 b_{8} + b_{9}) 
\\ & \nonumber
- a_{1} (24 b_{10} + 2 b_{11} + 6 b_{2} + 2 b_{3} + b_{5} + b_{6} + 3 b_{7} + 6 b_{8} + 7 b_{9})
\end{align}
\begin{align}
c_{7} &= 2 \big(a_{1} (4 b_{3} b_{4} + 2 b_{3} b_{5} + 2 b_{4} b_{5} + b_{5}^2 + 2 b_{4} b_{6} + b_{5} b_{6} + 8 b_{3} b_{7} + 3 b_{5} b_{7} + 2 b_{11} (-2 b_{2} 
\\ & \nonumber
+ 2 b_{4} + b_{5} + 3 b_{7} - 2 b_{8}) - 2 b_{5} b_{8} - 8 b_{6} b_{8} - 2 b_{7} b_{8} - 4 b_{8}^2 - 2 b_{10} (6 b_{11} - 2 b_{3} + 3 b_{5} 
\nonumber \\ & 
+ 15 b_{6} + 8 b_{8}) + (2 b_{3} + 8 b_{4} + 4 b_{5} - 3 b_{6} + 15 b_{7} - 4 b_{8}) b_{9} + 3 b_{9}^2 - 2 b_{2} (b_{5} + 4 b_{6} + 2 b_{8} 
\nonumber \\ & 
+ b_{9})) + a_{2} (4 b_{2}^2 + 4 b_{2} b_{3} - (2 b_{4} + b_{5})^2 + 2 b_{2} b_{6} - 2 b_{4} b_{6} - b_{5} b_{6} + 2 b_{10} (8 b_{2} + 8 b_{3} - 2 b_{4} 
\nonumber \\ & 
- b_{5} + 3 b_{6}) + 4 b_{2} b_{7} + 2 b_{3} b_{7} - 16 b_{4} b_{7} - 8 b_{5} b_{7} - 3 b_{6} b_{7} - 15 b_{7}^2 + 4 b_{2} b_{8} + 4 b_{3} b_{8} 
\nonumber \\ & 
+ 2 b_{6} b_{8} + 2 b_{7} b_{8} + (2 b_{2} + 2 b_{3} - 2 b_{4} - b_{5} - 3 b_{7}) b_{9} 
\nonumber \\ & 
+ 2 b_{1} (30 b_{10} + 8 b_{2} - 2 b_{4} - b_{5} + 8 b_{8} + 3 b_{9}))\big) 
\end{align}
\begin{align}
c_{8} &= 2\big(b_{1} + 16 b_{10} + b_{11} + 4 b_{2} + b_{3} + b_{4} + b_{5} + b_{6} + 4 (b_{7} + b_{8} + b_{9})\big) 
\end{align}
\begin{align}
c_{9} &= -2\big(4 b_{11} b_{2} + 4 b_{2}^2 + 4 b_{2} b_{3} + 12 b_{11} b_{4} + 4 b_{2} b_{4} + 12 b_{3} b_{4} + 4 b_{11} b_{5} - 4 b_{2} b_{5}
\\ & 
+ 4 b_{3} b_{5} + 4 b_{4} b_{5} + b_{5}^2 - 20 b_{2} b_{6} - 2 b_{5} b_{6} - 3 b_{6}^2 + 24 b_{11} b_{7} + 4 b_{2} b_{7} + 24 b_{3} b_{7} + 4 b_{5} b_{7} 
\nonumber \\ & 
- 12 b_{6} b_{7} - 3 b_{7}^2 + 4 b_{10} (3 b_{11} + 4 b_{2} + 3 b_{3} + 3 b_{4} - 5 b_{5} - 21 b_{6} - 12 b_{8}) - 8 b_{2} b_{8} - 8 b_{5} b_{8} 
\nonumber \\ & 
- 24 b_{6} b_{8} - 12 b_{7} b_{8} - 12 b_{8}^2 + 2 (2 b_{2} + 12 b_{4} + 2 b_{5} - 6 b_{6} + 21 b_{7} - 6 b_{8}) b_{9} - 3 b_{9}^2 
\nonumber \\ & \nonumber
+ 4 b_{1} (3 b_{10} + 3 b_{11} + b_{2} + 3 b_{3} + b_{5} + 6 b_{9})\big) 
\end{align}
\begin{align}
c_{10} &= -4\big(16 b_{11} b_{2}^2 + 4 b_{2}^3 + 12 b_{11} b_{2} b_{3} + 4 b_{2}^2 b_{3} - 4 b_{11} b_{3}^2 - 4 b_{2} b_{3}^2 - 4 b_{3}^3 + 4 b_{2}^2 b_{4} 
\\ & 
+ 4 b_{11} b_{3} b_{4} + 4 b_{2} b_{3} b_{4} - 16 b_{11} b_{4}^2 - 4 b_{2} b_{4}^2 - 4 b_{4}^3 - 4 b_{11} b_{2} b_{5} - 4 b_{2} b_{3} b_{5} - 4 b_{3}^2 b_{5} 
\nonumber \\ & 
- 12 b_{11} b_{4} b_{5} - 4 b_{2} b_{4} b_{5} + 4 b_{3} b_{4} b_{5} - 4 b_{4}^2 b_{5} - 2 b_{11} b_{5}^2 - 3 b_{2} b_{5}^2 + b_{3} b_{5}^2 + b_{4} b_{5}^2 + b_{5}^3 
\nonumber \\ & 
+ 4 b_{2}^2 b_{6} - 4 b_{3}^2 b_{6} + 16 b_{3} b_{4} b_{6} - 4 b_{4}^2 b_{6} - 10 b_{2} b_{5} b_{6} + 6 b_{3} b_{5} b_{6} + 6 b_{4} b_{5} b_{6} + 4 b_{5}^2 b_{6} 
\nonumber \\ & 
- 15 b_{2} b_{6}^2 + 3 b_{3} b_{6}^2 + 3 b_{4} b_{6}^2 + 3 b_{5} b_{6}^2 + 12 b_{11} b_{2} b_{7} + 4 b_{2}^2 b_{7} + 12 b_{11} b_{3} b_{7} - 4 b_{3}^2 b_{7} 
\nonumber \\ & 
- 60 b_{11} b_{4} b_{7} - 12 b_{2} b_{4} b_{7} + 4 b_{3} b_{4} b_{7} - 16 b_{4}^2 b_{7} - 24 b_{11} b_{5} b_{7} - 10 b_{2} b_{5} b_{7} + 6 b_{3} b_{5} b_{7} 
\nonumber \\ & 
- 6 b_{4} b_{5} b_{7} + 4 b_{5}^2 b_{7} - 6 b_{2} b_{6} b_{7} + 30 b_{3} b_{6} b_{7} - 6 b_{4} b_{6} b_{7} + 12 b_{5} b_{6} b_{7} - 54 b_{11} b_{7}^2 - 15 b_{2} b_{7}^2 
\nonumber \\ & 
+ 3 b_{3} b_{7}^2 - 15 b_{4} b_{7}^2 + 3 b_{5} b_{7}^2 + 12 b_{11} b_{2} b_{8} + 4 b_{2}^2 b_{8} + 16 b_{11} b_{3} b_{8} + 4 b_{2} b_{3} b_{8} + 4 b_{11} b_{4} b_{8} 
\nonumber \\ & 
+ 4 b_{2} b_{4} b_{8} + 12 b_{3} b_{4} b_{8} - 4 b_{2} b_{5} b_{8} + 4 b_{3} b_{5} b_{8} + 4 b_{4} b_{5} b_{8} + b_{5}^2 b_{8} - 12 b_{2} b_{6} b_{8} 
\nonumber \\ & 
+ 4 b_{3} b_{6} b_{8} + 4 b_{4} b_{6} b_{8} - 6 b_{5} b_{6} b_{8} - 15 b_{6}^2 b_{8} + 12 b_{11} b_{7} b_{8} + 16 b_{3} b_{7} b_{8} + 4 b_{4} b_{7} b_{8}
\nonumber \\ & 
+ 6 b_{5} b_{7} b_{8} - 6 b_{6} b_{7} b_{8} + 3 b_{7}^2 b_{8} - 4 b_{11} b_{8}^2 - 4 b_{2} b_{8}^2 - 4 b_{5} b_{8}^2 - 16 b_{6} b_{8}^2 - 4 b_{7} b_{8}^2 - 4 b_{8}^3 
\nonumber \\ & 
+ 2 b_{10} (8 b_{2}^2 - 2 b_{3}^2 + 8 b_{3} b_{4} - 2 b_{4}^2 - b_{5}^2 + 6 b_{3} b_{6} + 6 b_{4} b_{6} - 12 b_{5} b_{6} - 27 b_{6}^2 
\nonumber \\ & 
+ b_{2} (6 b_{3} + 6 b_{4} - 2 b_{5} + 6 b_{6}) + 6 b_{11} (5 b_{2} + 5 b_{3} - b_{4} - b_{5} - b_{8}) + 2 (b_{3} + b_{4} - 3 (b_{5} + 5 b_{6})) b_{8} - 8 b_{8}^2) 
\nonumber \\ & 
+ 2 (2 b_{2}^2 - 8 b_{3}^2 - 2 b_{4}^2 + 3 b_{4} b_{5} + 2 b_{5}^2 + 15 b_{4} b_{6} + 6 b_{5} b_{6} - 3 b_{4} b_{7} + 6 b_{5} b_{7} + 27 b_{6} b_{7} 
\nonumber \\ & 
- b_{2} (6 b_{3} + 5 b_{5} + 3 (b_{6} + b_{7})) + (8 b_{4} + 3 b_{5} - 3 b_{6} + 15 b_{7}) b_{8} - 2 b_{8}^2 + b_{3} (2 b_{4} 
\nonumber \\ & 
- 3 (b_{5} + b_{6} + b_{7}) + 2 b_{8})) b_{9} + 3 (-5 b_{2} - 5 b_{3} + b_{4} + b_{5} + b_{8}) b_{9}^2 
\nonumber \\ & 
+ 2 b_{1} (8 b_{2}^2 - 8 b_{3}^2 + 6 b_{2} b_{4} + 2 b_{3} b_{4} - 2 b_{4}^2 - 2 b_{2} b_{5} - 6 b_{3} b_{5} - b_{5}^2 + 6 b_{10} (18 b_{11} + 5 b_{2} - b_{3} 
\nonumber \\ & 
+ 5 b_{4} - b_{5} - b_{8}) + 6 b_{2} b_{8} + 2 b_{3} b_{8} + 8 b_{4} b_{8} - 2 b_{8}^2 + 6 b_{11} (5 b_{2} - b_{3} - b_{4} - b_{5} + 5 b_{8}) 
\nonumber \\ & \nonumber
+ 6 (b_{2} - 5 b_{3} + b_{4} - 2 b_{5} + b_{8}) b_{9} - 27 b_{9}^2)\big) 
\end{align}

\pagebreak

\section{Definition of coupling constants $d_i$}

\label{App2}

Below we give the definitions of $d_i$ in \eqref{K}

\begin{align}&
d_1 = (a_1 + a_2)^2
\end{align}
\begin{align}&
d_2 = -4 (a_2^2 (3 b_1 + 3 b_{10} + b_2 + 3 b_4 + b_5 + 6 b_7) + a_1^2 (3 b_{10} + 3 b_{11} + b_2 + 3 b_3 + b_5 + 6 b_9) 
\\ & \nonumber 
- a_1 a_2 (42 b_{10} + 10 b_2 + b_5 + 3 b_6 + 6 (b_7 + 2 b_8 + b_9)))
\end{align}
\begin{align} 
d_3 = & -4 (a_2^2 (8 b_2^2 + 6 b_2 b_3 - 2 b_3^2 + 2 b_3 b_4 - 8 b_4^2 - 2 b_2 b_5 - 6 b_4 b_5 - b_5^2 + 6 b_2 b_7 + 6 b_3 b_7 - 30 b_4 b_7 
\\ & \nonumber 
- 12 b_5 b_7 - 27 b_7^2 + 6 b_{10} (5 b_2 + 5 b_3 - b_4 - b_5 - b_8) + 2 (3 b_2 + 4 b_3 + b_4 + 3 b_7) b_8 - 2 b_8^2 
\\ & \nonumber 
+ 6 b_1 (18 b_{10} + 5 b_2 - b_3 - b_4 - b_5 + 5 b_8)) + a_1^2 (8 b_2^2 - 8 b_3^2 + 6 b_2 b_4 + 2 b_3 b_4 - 2 b_4^2 - 2 b_2 b_5 - 6 b_3 b_5 
\\ & \nonumber 
- b_5^2 + 6 b_{10} (18 b_{11} + 5 b_2 - b_3 + 5 b_4 - b_5 - b_8) + 6 b_2 b_8 + 2 b_3 b_8 + 8 b_4 b_8 - 2 b_8^2 + 6 b_{11} (5 b_2 - b_3 
\\ & \nonumber 
- b_4 - b_5 + 5 b_8) + 6 (b_2 - 5 b_3 + b_4 - 2 b_5 + b_8) b_9 - 27 b_9^2) + 2 a_1 a_2 (2 b_2^2 - 2 (b_3^2 - 4 b_3 b_4 + b_4^2) 
\\ & \nonumber 
- 5 b_2 b_5 + 3 b_3 b_5 + 3 b_4 b_5 + 2 b_5^2 + 3 b_3 b_6 + 3 b_4 b_6 + 3 b_5 b_6 + 15 b_3 b_7 - 3 b_4 b_7 + 6 b_5 b_7 + 6 b_{10} (b_2 
\\ & \nonumber 
+ b_3 + b_4 - 2 b_5 - 9 b_6 - 5 b_8) + 2 b_3 b_8 + 2 b_4 b_8 - 3 b_5 b_8 - 15 b_6 b_8 - 3 b_7 b_8 - 8 b_8^2 - 3 (b_3 - 5 b_4 
\\ & \nonumber 
- 2 b_5 - 9 b_7 + b_8) b_9 - 3 b_2 (5 b_6 + b_7 + 2 b_8 + b_9)))
\end{align}
\begin{align}
d_4 = & - 4 (b_1 + 16 b_{10} + b_{11} + 4 b_2 + b_3 + b_4 + b_5 + b_6 + 4 (b_7 + b_8 + b_9))
\end{align}
\begin{align}
d_5 = & 4 (4 b_{11} b_2 + 4 b_2^2 + 4 b_2 b_3 + 12 b_{11} b_4 + 4 b_2 b_4 + 12 b_3 b_4 + 4 b_{11} b_5 - 4 b_2 b_5 + 4 b_3 b_5 + 4 b_4 b_5 + b_5^2 
\\ & \nonumber 
- 20 b_2 b_6 - 2 b_5 b_6 - 3 b_6^2 + 24 b_{11} b_7 + 4 b_2 b_7 + 24 b_3 b_7 + 4 b_5 b_7 - 12 b_6 b_7 - 3 b_7^2 + 4 b_{10} (3 b_{11} 
\\ & \nonumber 
+ 4 b_2 + 3 b_3 + 3 b_4 - 5 b_5 - 21 b_6 - 12 b_8) - 8 b_2 b_8 - 8 b_5 b_8 - 24 b_6 b_8 - 12 b_7 b_8 - 12 b_8^2 + 2 (2 b_2 
\\ & \nonumber 
+ 12 b_4 + 2 b_5 - 6 b_6 + 21 b_7 - 6 b_8) b_9 - 3 b_9^2 + 4 b_1 (3 b_{10} + 3 b_{11} + b_2 + 3 b_3 + b_5 + 6 b_9))
\end{align}
\begin{align}
d_6 = & 8 (16 b_{11} b_2^2 + 4 b_2^3 + 12 b_{11} b_2 b_3 + 4 b_2^2 b_3 - 4 b_{11} b_3^2 - 4 b_2 b_3^2 - 4 b_3^3 + 4 b_2^2 b_4 + 4 b_{11} b_3 b_4 + 4 b_2 b_3 b_4 
\\ & \nonumber 
- 16 b_{11} b_4^2 - 4 b_2 b_4^2 - 4 b_4^3 - 4 b_{11} b_2 b_5 - 4 b_2 b_3 b_5 - 4 b_3^2 b_5 - 12 b_{11} b_4 b_5 - 4 b_2 b_4 b_5 + 4 b_3 b_4 b_5 
\\ & \nonumber 
- 4 b_4^2 b_5 - 2 b_{11} b_5^2 - 3 b_2 b_5^2 + b_3 b_5^2 + b_4 b_5^2 + b_5^3 + 4 b_2^2 b_6 - 4 b_3^2 b_6 + 16 b_3 b_4 b_6 - 4 b_4^2 b_6 - 10 b_2 b_5 b_6 
\\ & \nonumber 
+ 6 b_3 b_5 b_6 + 6 b_4 b_5 b_6 + 4 b_5^2 b_6 - 15 b_2 b_6^2 + 3 b_3 b_6^2 + 3 b_4 b_6^2 + 3 b_5 b_6^2 + 12 b_{11} b_2 b_7 + 4 b_2^2 b_7 + 12 b_{11} b_3 b_7 
\\ & \nonumber 
- 4 b_3^2 b_7 - 60 b_{11} b_4 b_7 - 12 b_2 b_4 b_7 + 4 b_3 b_4 b_7 - 16 b_4^2 b_7 - 24 b_{11} b_5 b_7 - 10 b_2 b_5 b_7 + 6 b_3 b_5 b_7 
\\ & \nonumber 
- 6 b_4 b_5 b_7 + 4 b_5^2 b_7 - 6 b_2 b_6 b_7 + 30 b_3 b_6 b_7 - 6 b_4 b_6 b_7 + 12 b_5 b_6 b_7 - 54 b_{11} b_7^2 - 15 b_2 b_7^2 + 3 b_3 b_7^2 
\\ & \nonumber 
- 15 b_4 b_7^2 + 3 b_5 b_7^2 + 12 b_{11} b_2 b_8 + 4 b_2^2 b_8 + 16 b_{11} b_3 b_8 + 4 b_2 b_3 b_8 + 4 b_{11} b_4 b_8 + 4 b_2 b_4 b_8 + 12 b_3 b_4 b_8 
\\ & \nonumber 
- 4 b_2 b_5 b_8 + 4 b_3 b_5 b_8 + 4 b_4 b_5 b_8 + b_5^2 b_8 - 12 b_2 b_6 b_8 + 4 b_3 b_6 b_8 + 4 b_4 b_6 b_8 - 6 b_5 b_6 b_8 - 15 b_6^2 b_8 
\\ & \nonumber 
+ 12 b_{11} b_7 b_8 + 16 b_3 b_7 b_8 + 4 b_4 b_7 b_8 + 6 b_5 b_7 b_8 - 6 b_6 b_7 b_8 + 3 b_7^2 b_8 - 4 b_{11} b_8^2 - 4 b_2 b_8^2 - 4 b_5 b_8^2 
\\ & \nonumber 
- 16 b_6 b_8^2 - 4 b_7 b_8^2 - 4 b_8^3 + 2 b_{10} (8 b_2^2 - 2 b_3^2 + 8 b_3 b_4 - 2 b_4^2 - b_5^2 + 6 b_3 b_6 + 6 b_4 b_6 - 12 b_5 b_6 
\\ & \nonumber 
- 27 b_6^2 + b_2 (6 b_3 + 6 b_4 - 2 b_5 + 6 b_6) + 6 b_{11} (5 b_2 + 5 b_3 - b_4 - b_5 - b_8) + 2 (b_3 + b_4 - 3 (b_5 + 5 b_6)) b_8 
\\ & \nonumber 
- 8 b_8^2) + 2 (2 b_2^2 - 8 b_3^2 - 2 b_4^2 + 3 b_4 b_5 + 2 b_5^2 + 15 b_4 b_6 + 6 b_5 b_6 - 3 b_4 b_7 + 6 b_5 b_7 + 27 b_6 b_7 - b_2 (6 b_3 
\\ & \nonumber 
+ 5 b_5 + 3 (b_6 + b_7)) + (8 b_4 + 3 b_5 - 3 b_6 + 15 b_7) b_8 - 2 b_8^2 + b_3 (2 b_4 - 3 (b_5 + b_6 + b_7) + 2 b_8)) b_9 
\\ & \nonumber 
+ 3 (-5 b_2 - 5 b_3 + b_4 + b_5 + b_8) b_9^2 + 2 b_1 (8 b_2^2 - 8 b_3^2 + 6 b_2 b_4 + 2 b_3 b_4 - 2 b_4^2 - 2 b_2 b_5 - 6 b_3 b_5 - b_5^2 
\\ & \nonumber 
+ 6 b_{10} (18 b_{11} + 5 b_2 - b_3 + 5 b_4 - b_5 - b_8) + 6 b_2 b_8 + 2 b_3 b_8 + 8 b_4 b_8 - 2 b_8^2 + 6 b_{11} (5 b_2 - b_3 - b_4 - b_5 + 5 b_8) 
\\ & \nonumber 
+ 6 (b_2 - 5 b_3 + b_4 - 2 b_5 + b_8) b_9 - 27 b_9^2))
\end{align}

\pagebreak

\bibliographystyle{JHEP}
\bibliography{gb}

\end{document}